\documentclass[12pt]{article}
\usepackage{draft,graphicx}
\usepackage{dcolumn}
\usepackage{bm,tikz-cd}

\def\ma{\mathcal}
\def\ie{\begin{equation}\begin{aligned}}
\def\fe{\end{aligned}\end{equation}}

\begin{document}

\title{The bi-adjoint scalar $\ell$-loop planar integrand recursion and graded inverse variables}

\author{Yi-Xiao Tao}
\email{taoyx21@mails.tsinghua.edu.cn}
\address{Department of Mathematical Sciences, Tsinghua University, Beijing 100084, China\\
Nordita, KTH Royal Institute of Technology and Stockholm University, Hannes Alf\'{v}ens v\"{a}g 12, SE-106 91 Stockholm, Sweden
}%

\begin{abstract}
Previously in \cite{Tao:2025fch}, we constructed the $\ell$-loop planar integrands using loop components and loop kernels by some recursion rules. In this paper, we propose a new formalism to express the loop kernel recursion. We define ``graded inverse variables" to make the loop kernel recursion more elegant. And the graph factor, including the symmetry factor, can be figured out from each monomial of some variables. This new formalism makes the previous $\ell$-loop integrand recursion clearer.
\end{abstract}
\newpage
\section{Introduction}
Off-shell methods \cite{Schubert:2001he} in amplitudes have a long history\footnote{In contrast, for on-shell methods, see \cite{Elvang:2013cua,Travaglini:2022uwo} for some reviews.}. Off-shell methods can reveal a richer structure than the on-shell methods and can help us understand more about the theory. In addition, off-shell objects are very useful in generating loop integrands or computing amplitudes recursively. One of the most successful off-shell objects is the Berends-Giele (BG) current \cite{Berends:1987me}. The BG current has only one leg off-shell and all other legs on-shell; hence, it is partly simplified by the on-shell legs but can still show some hidden information of the theory via the off-shell leg. Due to its good properties, the BG current has many applications in different cases, including scattering amplitudes, (A)dS correlators, celestial amplitudes, and so on \cite{Mizera:2018jbh,Gomez:2021shh,Armstrong:2022mfr,Chattopadhyay:2021udc,Chattopadhyay:2024kdq,Wu:2021exa,Mafra:2010jq,Mafra:2015vca,Mafra:2016ltu,Tao:2023wls,Tao:2023yxy,Tao:2024vcz,Frost:2020eoa,Cho:2021nim,Chen:2023bji,Tao:2022nqc}. However, the case that more legs being off-shell is still lack of study, which means that more off-shell methods still need to be developed. 

Multi-particle solutions of the equation of motion can be regarded as an off-shell generalization of BG currents, where all external legs are off-shell. There are some examples of obtaining 1-loop integrands from the multi-particle solution of the classical equation of motion \cite{Gomez:2022dzk}. Actually, we can also obtain the multi-particle solution of a quantum version of the equation of motion directly \cite{Lee:2022aiu,Garg:2024icm}. Multi-particle solutions are more suitable for dealing with the loop integrands. We do not need to worry about the missing information because of the on-shell condition, since all external legs of the multi-particle solutions are off-shell.

Recently, we proposed a new off-shell method, which can help us to obtain $\ell$-loop planar integrands recursively in \cite{Tao:2025fch}. In the colored theory, planar integrands correspond to the partial amplitudes associated with a single-trace structure in the color-trace decomposition \cite{Elvang:2013cua}. This method starts with the equation of motion of the bi-adjoint theory, and finally reaches the whole $\ell$-loop planar integrand results. This systematic approach to $\ell$-loop planar integrands is based on the comb component of the multi-particle solutions and the $\ell$-loop kernel. These two objects will be defined later. Based on these two objects, together with generalized BG currents, we will obtain the $\ell$-loop planar integrands recursively\footnote{The result for the Yang-Mills theory can also be found in \cite{Arkani-Hamed:2024tzl,Cao:2025mlt}}. This method can be generalized to any theory to obtain the $\ell$-loop planar integrands by considering the planar component of the multi-particle solution to rule out the non-planar part\footnote{In this paper, the word `` $\ell$-loop integrands" is equivalent to ``$\ell$-loop planar integrands" since we will not discuss the non-planar ones.}. In this paper, we will define new variables called ``graded inverse variables" to reconstruct this recursion\footnote{Some results based on inverse variables can be found in \cite{Arkani-Hamed:2024pzc}}. The recursion based on such new variables has a more elegant formalism and improves some shortcomings of the recursion in the previous formalism.

This paper is organized as follows. We will review the recursion of $\ell$-loop planar integrands of the bi-adjoint scalar theory in section \ref{sec2}. In section \ref{sec3}, we will define the ``graded inverse variables" and show how to express our loop kernel recursion using these variables.

\section{A review of the off-shell engineering}\label{sec2}
In this section, we will review how to construct $\ell$-loop integrands from the equation of motion, which is also named as ``off-shell engineering", using bi-adjoint scalar theory as an example, since we will only discuss the bi-adjoint scalar theory in this paper. For more comments, see \cite{Tao:2025fch}.
\subsection{Comb components and 1-loop kernels}
Consider the bi-adjoint scalar theory with the following equation of motion:
\ie
\square \phi=\frac{1}{2}[[\phi,\phi]].
\fe
where $\phi=\phi_{a\tilde{a}} T^a\otimes\tilde{T}^{\tilde{a}}$. The notation $[[]]$ denotes the double commutator of two different generator $T^a$ and $\tilde{T}^{\tilde{a}}$, i.e.
\ie
[[T^a\otimes\tilde{T}^{\tilde{a}},T^b\otimes\tilde{T}^{\tilde{b}}]]=[T^a,T^b]\otimes[\tilde{T}^{\tilde{a}},\tilde{T}^{\tilde{b}}]
\fe
Using the perturbiner method \cite{Selivanov:1997aq,Selivanov:1998hn,Rosly:1997ap,Rosly:1998vm,Mizera:2018jbh}, we can obtain the multi-particle solution of the equation of motion above:
\ie
\phi&=\sum_{P,Q}\phi_{P|Q}e^{ik_P\cdot x}T^{a_P}\otimes\tilde{T}^{\tilde{a}_Q}\\
s_P\phi_{P|Q}&=\sum_{P=XY}\sum_{Q=WZ}(\phi_{X|W}\phi_{Y|Z}-\phi_{X|Z}\phi_{Y|W})
\fe
with $P=p_1\cdots p_m$, $T^{a_P}=T^{a_1}T^{a_2}\cdots T^{a_m}$, and $s_P=k_P^2=(\sum_{i=1}^mk_{p_i})^2$. Here we will not impose any on-shell conditions, which means that $k_i^2\neq0$ for $i\in P$. Now we define the comb component of the multi-particle solution, which is one of the most important objects in our recursion of loop integrands:
\ie
s_P\phi^{\rm comb}_{12\cdots m|Q}=\sum_{Q=WZ}(\phi^{\rm comb}_{12\cdots m-1|W}\phi_{m|Z}-(W\leftrightarrow Z)).
\fe
For simplicity, here we set $P=12\cdots m$. This recursive definition starts with $\phi^{\rm comb}_{i|i}=\phi_{i|i}$ and we impose $Q=P$ in this paper. We can write down a closed formula for the comb component:
\ie
\phi^{\rm comb}_{12\cdots m|12\cdots m}=\frac{\prod_{i=1}^m\phi_{i|i}}{s_{1\cdots m}s_{1\cdots m-1}\cdots s_{12}}
\fe
We need to emphasize that we will not use $\phi_{p|q}=\delta_{pq}$, which corresponds to single-particle states, since all external legs are off-shell.

We can define the $m$-way 1-loop kernel using the comb component, as the starting point of the recursion of the loop integrands. We first consider the comb component $\phi^{\rm comb}_{l12\cdots m|l12\cdots m}$, then using the sewing procedure \cite{Gomez:2022dzk} $\phi_{l|l}\to1/l_1^2$ we can obtain the $m$-way 1-loop kernel
\ie
I^{\rm kernel}_{1,m}=\frac{\prod_{i=1}^m\phi_{i|i}(1-\frac{1}{2}\delta_{m,2})}{l_1^2(l_1+k_1)^2\cdots(l_1+\sum_{i=1}^{m-1}k_i)^2}
\fe
with the loop momentum $l_1$. Note that when $m=2$ we have a symmetry factor of 1/2. Consider an $n$-point 1-loop integrand $I^{\rm 1-loop}(12\cdots n|Q)$, we just need to divide $(12\cdots n)$ into $m$ parts, say $P_m$. Note that when constructing loop integrands, the ordered sets here are cyclic sets, which means they are equivalent under cyclic permutation. As an example, the division we considered here also includes cases like $(n-1,n,1,2|3,4,\cdots,n-2)$. Then we do the following replacement
\ie\label{replace}
I^{\rm 1-loop}_m(12\cdots n|Q)=I^{\rm kernel}_{1,m}\bigg|_{k_i\to k_{P_i},\phi_{i|i}\to\Phi_{P_i|Q_i}}.
\fe
Here, $\{Q_i\}$ is a division of the cyclic set $Q$, i.e. $(Q_1|Q_2|\cdots|Q_m)$ is a division of $Q$ up to a cyclic permutation of $Q$. And $\Phi_{P_i|Q_i}$ here is the Berends-Giele current with all legs in $P_i$ on-shell comparing to the off-shell multi-particle solution $\phi_{P_i|Q_i}$, which means that in $\Phi_{P_i|Q_i}$ we need to impose $\phi_{i|j}=\delta_{ij}$. Note that we must find a division so that for every $i$, the elements in $Q_i$ must be the same in $P_i$, otherwise this term of the integrand will vanish. We will show examples later to show the differences between different $Q$. The whole $n$-point 1-loop integrands can be expressed as
\ie\label{1-loop}
I^{\rm 1-loop}(12\cdots n|Q)=\sum_{m=2}^{n}\sum_{\text{ $m$-division of $P$}}I^{\rm kernel}_{1,m}\bigg|_{k_i\to k_{P_i}\atop \phi_{i|i}\to\Phi_{P_i|Q_i}}
\fe
When $m=1$, we will get a tadpole, which will be zero after integrating the loop momenta. Therefore, we start the sum with $m=2$. Some examples are shown in \cite{Tao:2025fch}.

\subsection{Graph factors: avoid overcounting}
Before we go to the $\ell$-loop case, we need to define the graph factor here in order to avoid overcounting. For an $\ell$-loop Feynman diagram, a graph factor can be written as $g=S\times \frac{1}{\ell-r}$. Here, $S$ is the symmetry factor coming from the symmetry of the Feynman diagram, and $r$ is defined as below. We first define the largest loop to be the loop that is connected with all external legs and no other loop outside it. Then we consider all loops that are independent in this largest loop, i.e., the regions of different loops will not overlap. After choosing the largest loop, we will ignore the external legs and the 3-point vertices connected to them and regard the diagram just as a vacuum diagram. Then, each loop will have some propagators in common with the largest loop. If there is 1 common propagator, we assign the loop a value of 0; otherwise 1. This value is called the loop value. Then $r$ is the sum of the values of all loops.

In the bi-adjoint scalar case, the symmetry factors in our recursion can be figured out easily. In this case, all symmetry factors come from the 2-way kernel parts of the Feynman diagrams. Every part contributes a factor of 1/2. This fact can be understood as follows. For a 2-way loop kernel, it has an automorphism: it is invariant if one turns it upside down. Note that the symmetry factor here is not the same as the traditional one that only comes from the Feynman rules. The symmetry factor here just aims to avoid the overcounting mentioned above, from both the Feynman rules and the recursion.

The loop kernel without the symmetry factor $S$ in each diagram is called a bare loop kernel. This concept is important when considering recursion, since when we go to higher-loop cases, the symmetry factor will change; hence, we only need to consider the symmetry factor in the last recursion. However, the factor $\frac{1}{\ell-r}$ here is to avoid overcounting in the following recursion; hence, we must consider it in every recursion.

\subsection{Towards $\ell$-loop integrands}
The general definition of the $\ell$-loop kernel is the sum of all the indivisible $\ell$-loop diagrams with all external legs off-shell. Here, by the word ``indivisible", we mean that these diagrams cannot be reduced to two parts by cutting a single propagator. The $\ell$-loop kernel can be obtained from the $(\ell-1)$-loop kernel recursively. For an $m_\ell$-way $\ell$-loop kernel, we have the following steps to construct it:
\begin{enumerate}
\item Consider ordered cyclic set $(b_\ell ,a_\ell,1,2, \cdots ,m_\ell-1,m_\ell)$ and the cyclic permutation of $(12\cdots m_n)$, i.e. there are $m_\ell$ sets in total. Note that due to the cyclic properties, $(1,2,\cdots,m_\ell-1,m_\ell,a_\ell,b_\ell)$ is equivalent to $(a_\ell,b_\ell,1,2\cdots (m_\ell-1)m_\ell)$. 
\item Then, for $k$-divisions, which means we divide a set into $k$ parts, of each set, we set: i) $b_{\ell}$ itself to be one part of the $k$-division, ii) For a part not involving $a_{\ell}$, there can only be 1 element in the part, like $(b_{\ell}|a_{\ell},1|2|3)$. There is only 1 case for a given $k$ and a given set. 
\item Consider a $k$-way bare $(\ell-1)$-loop kernel, replace $\phi_{i|i}$ of this kernel with the comb component $\phi^{\rm comb}$ of each part of a $k$-division, just similar to \eqref{replace}. Then replace $\phi_{a_\ell|a_\ell}\phi_{b_\ell|b_\ell}$ with $1/l_\ell^2$ and $k_{a_\ell}=-k_{b_\ell}=l_\ell$. This replacement is equivalent to turning legs $a_\ell$ and $b_\ell$ into an internal line. Then add the corresponding graph factor $g$ to each case. Since we have demonstrated the construction of the Feynman diagrams, one can easily draw diagrams according to the steps above and find the corresponding graph factors.
\item Sum over $k$-division for all sets and all $k\geq 2$, just like \eqref{1-loop}. Finally we will obtain a $m_\ell$-way $\ell$-loop kernel from $(\ell-1)$-loop kernel $I^{\rm kernel}_{\ell,m_\ell}$.
\end{enumerate}
Then, after defining the $\ell$-loop current
\ie
\Phi^{(\ell)}_{P|Q}=\frac{1}{k_P^2}I^{\rm \ell-loop}(Pl|Ql)\bigg|_{k_l\to -k_P},
\fe
we can obtain the total $\ell$-loop planar integrand:
\ie\label{n-loop}
I^{\rm \ell-loop}(12\cdots n|Q)=&\sum_{k=2}^{n}\sum_{\text{ $k$-division of $P$}}I^{\rm kernel}_{\ell,k}\bigg|_{k_i\to k_{P_i}\atop \phi_{i|i}\to\Phi_{P_i|Q_i}}\\
&+\frac{1}{\ell}(\sum_{m=1}^{\ell-1}\sum_{k=2}^{n}\sum_{\ell_i\atop\sum_{i=1}^k \ell_i=\ell-m}\sum_{\text{ $k$-division of $P$}}mI^{\rm kernel}_{m,k}\bigg|_{k_i\to k_{P_i}\atop \phi_{i|i}\to\Phi^{(\ell_i)}_{P_i|Q_i}})
\fe
with $P=123\cdots n$. Note that the factor $m/\ell$ in the second line is important to avoid overcounting.

\subsection{Question: how to obtain the graph factor without drawing diagrams?}
The recursion relation \eqref{n-loop} looks awesome. Nearly all steps can be followed conveniently after the inputs, namely the lower-loop objects, are given. However, there is still one thing that looks awkward: how to obtain the graph factor when we follow the recursion steps? It seems that we must draw diagrams at the same time as we follow these steps to find out these graph factors. Of course, following the recursive process, the corresponding diagrams are manifest. But can we do better? Can we find out the graph factors without drawing corresponding diagrams? 

The most natural idea is to define variables for each propagator since the structure of each diagram is described by propagators. Meanwhile, we also know the integrand of each diagram, which consists of some factors involving momenta. This allows us to construct a map between propagator variables and the integrands. In section \ref{sec3}, we will propose a new formalism of the recursion based on such new variables. We will rewrite the recursion step for the loop kernel more compactly and construct a map between the polynomials of these new variables and the integrands. In this new formalism, the graph factor will be given by the structure of the monomials of the new variables, which means that we do not need to draw diagrams anymore.

\section{A new formalism of the recursion}\label{sec3}
The recursion for the loop kernels above requires us to draw diagrams to figure out the graph factors, which means that we cannot obtain the loop integrands by only dealing with the algebra, but need to consider the corresponding Feynman diagram as auxiliary. The above problem not only causes extra work when we perform the recursion, but also presents coding difficulties. To solve this problem, we need to define suitable variables to represent each propagator and deal with the polynomials of these variables, as the graph factor problems are highly relevant to the loop topology, which can be determined by considering the structure of the propagators in the integrands.

In this section, we will propose a new formalism using the so-called ``graded inverse variables". In this formalism, we do not need to draw diagrams when we do the loop kernel recursion. Instead, all graph factors can be figured out from the structures of the monomials of graded inverse variables.

\subsection{Graded inverse variables}
The inverse variables $x_{kl}$ represent the line connecting two vertices $k$ and $l$. In our recursion, whether the vertices connect with an external leg is important; hence, we will use extra position labels $e$, which means ``external", and $i$, which means ``internal", to distinguish this thing. The corresponding vertices are also called external vertices or internal vertices. For example, the variable $x_{kl,ei}$ means a line between $k$ and $l$ with only $k$ connecting to an external leg. Note that $x_{kl,ei}=x_{lk,ie}$. Such variables can be mapped to the propagator between two vertices and some $\phi_{j|j}$s, which will be considered in detail in later discussions.

Graded inverse variables naturally appear in the loop kernel recursion. We add an extra grade label $\ell$ to mark the generated loop by adding these variables when we follow the recursion steps that will be given later, say $x_{12,ei}^{(\ell)}$. We will see some examples later.

Now we will introduce two structures, which are useful to figure out the graph factors: the largest loop and the 2-way kernel structures. A loop with $n$ vertices in new variables is defined as
\ie
x_{k_1k_2}x_{k_2k_3}\cdots x_{k_{n-1}k_{n}}x_{k_nk_1}.
\fe
Here we do not care about the position labels and the graded labels. The largest loop of a monomial of graded inverse variables is defined as follows:
\begin{enumerate}
    \item It is the loop involving all vertices with position label $e$.
    \item When several choices satisfy the condition above, choose the one with higher grade variables. For example, if there are two ways $x_{13}^{(3)}$ and $x_{12}^{(2)}x_{23}^{(2)}$ connecting vertices 1 and 3 in the loop, we should choose the former one as a part of the largest loop.
    \item  When several choices satisfy the condition above but with the same grade variables, choose the one with the fewest variables. For example, if there are two ways $x_{13}^{(2)}$ and $x_{12}^{(2)}x_{23}^{(1)}$ connecting vertices 1 and 3 in the loop, we should choose the former one as a part of the largest loop.
\end{enumerate}
We also need to define an operation to apply the largest loop we defined above in the loop kernel recursion. A contraction is defined as follows:
\ie\label{cont}
x_{jk,ae}^{(\ell)}x_{kl,eb}^{(\ell)}\to x_{jl,ab}^{(\ell)}
\fe
with $a,\ b=\{e,i\}$.

Then we need to define 2-way kernel structures. 2-way kernel structures are the 2-way kernel parts of a given Feynman diagram. These parts are related to the symmetry factors. A 2-way kernel structure labeled by $(k,\ l)$ defined as follows:
\begin{enumerate}
    \item It is a factor of the variable monomial with the form $x_{km,ai}x_{kn,ai}\cdots x_{pl,ib}x_{ql,ib}$. The grade label is not important here, and $a,\ b=\{e,i\}$.
    \item In this factor, all the vertices must be assigned a position label $i$ except the vertices $k$ and $l$.
    \item In this factor, all the vertex labels appear 3 times except the vertices $k$ and $l$.
    \item Specifically, the factor $x_{kl}x_{lk}$ is also a 2-way kernel structure.
\end{enumerate}
Note that in a given loop kernel, there can exist many 2-way kernel structures.

In the following subsections, we will use these graded inverse variables to construct the recursion process.

\subsection{Graph factors}
The graph factors of each monomial can be determined as follows. We first figure out the largest loop and totally contract it using \eqref{cont}, and then totally contract the whole monomial. After comparing the factors of each grade on both sides, the number of common variables will give us 0 or 1 in each grade: 0 for only one common variable and 1 for others. The sum of these 0‘s and 1’s of all grades is the factor $r$ mentioned in section \ref{sec2}. For later convenience, we define a linear operator $\ma{R}$:
\ie
\ma{R}(M_{\ell})=\frac{1}{\ell-r}M_{\ell}
\fe
where $M_{\ell}$ is a monomial of graded inverse variables with the highest grade $\ell$, and $r$ is the corresponding factor for $M_{\ell}$ as mentioned above.

The symmetry factor $S$ can be found by counting the number of 2-way kernel structures. Each 2-way kernel structure will contribute a factor of 1/2.

\subsection{Recursion using new variables}
To simplify the notation, we will use natural numbers to label the vertices. The number of ways of a loop kernel can be figured out by counting the number of $e$ position labels. A $(m+1)$-point $\ell$-grade comb component is
\ie
{\rm comb}^{(\ell)}(1,2,\cdots,m)=x^{(\ell)}_{12,ee}\cdots x^{(\ell)}_{(m-1)m,ee}.
\fe
Specifically,
\ie
{\rm comb}^{(\ell)}(1)=1.
\fe
Consequently, we will get
\ie
P^{\rm kernel}_{1,m}=x^{(1)}_{12,ee}\cdots x^{(1)}_{(m-1)m,ee}x^{(1)}_{m1,ee}.
\fe
To distinguish the integrands and the polynomials, we use another notation $P$ to denote the polynomials of graded inverse variables. Now we define the graded sewing operator, which is a linear operator, between 2 adjacent vertices $k_1,\ k_2$ and an ordered set $\{l_1,l_2,\cdots,l_n\}$:
\ie\label{sew}
\ma{S}_{k_1,k_2}^{(\ell_2)}(x_{jk_1,ae}^{(\ell_1)}x_{k_2m,eb}^{(\ell_1)}P,\{l_1,l_2,\cdots,l_n\})=x_{jk_1,ai}^{(\ell_1)}x_{k_2m,ib}^{(\ell_1)}x_{k_1l_1,ie}^{(\ell_2)}x_{k_2l_n,ie}^{(\ell_2)}{\rm comb}^{(\ell_2)}(l_1,l_2,\cdots,l_n)P\bigg|_{x_{k_1j',ea'}\to x_{k_1j',ia'}\atop x_{k_2m',eb'}\to x_{k_2m',ib'}}
\fe
where $P$ is the product of some graded inverse variables, and the vertices $j'$ and $m'$ are arbitrary. Specifically, we define
\ie
\ma{S}_{k_1,k_2}^{(\ell_2)}(x_{jk_1,ae}^{(\ell_1)}x_{k_2m,eb}^{(\ell_1)}P,\emptyset)=x_{jk_1,ai}^{(\ell_1)}x_{k_2m,ib}^{(\ell_1)}x_{k_1k_2,ii}^{(\ell_2)}P\bigg|_{x_{k_1j',ea'}\to x_{k_1j',ia'}\atop x_{k_2m',eb'}\to x_{k_2m',ib'}}
\fe
and
\ie
\ma{S}_{k_1,k_2}^{(\ell_2)}(x_{jk_1,ae}^{(\ell_1)}x_{k_2m,eb}^{(\ell_1)}P,\{l\})=x_{jk_1,ai}^{(\ell_1)}x_{k_2m,ib}^{(\ell_1)}x_{k_1l,ie}^{(\ell_2)}x_{k_2l,ie}^{(\ell_2)}P\bigg|_{x_{k_1j',ea'}\to x_{k_1j',ia'}\atop x_{k_2m',eb'}\to x_{k_2m',ib'}}
\fe
Then for a $m_\ell$-way $\ell$-loop kernel polynomial, we can obtain it recursively by the following steps
\begin{enumerate}
\item Consider set $(1,2, \cdots ,m_\ell-1,m_\ell)$ and its cyclic permutations. 
\item Then, for the $(k-1)$-division, there is only one part of the $(k-1)$-division that can have any number of elements, including 0; we set this part to be the first part of the division. The other parts must have only 1 element in it. For example, for the set $(1,2, \cdots,m_\ell-1,m_\ell)$, the only valid 2-division is $(1,2, \cdots,m_\ell-1|m_\ell)$.
\item Consider a $k$-way bare $(\ell-1)$-loop kernel polynomial, we sew the first part of the $(k-1)$-division to the first 2 external vertices of this $(\ell-1)$-loop kernel polynomial using the $\ell$-grade sewing operation \eqref{sew}. Then replace the other external vertices' labels with the other part of the $(k-1)$-division in order. We also need to add the graph factors for each monomial.
\item Sum over all possible sets and all $k\geq2$, we will obtain a $m_\ell$-way $\ell$-loop kernel polynomial $P_{\ell,m_{\ell}}^{\rm kernel}$.
\end{enumerate}

\begin{figure}
	\centering
    \includegraphics[width=0.80\textwidth]{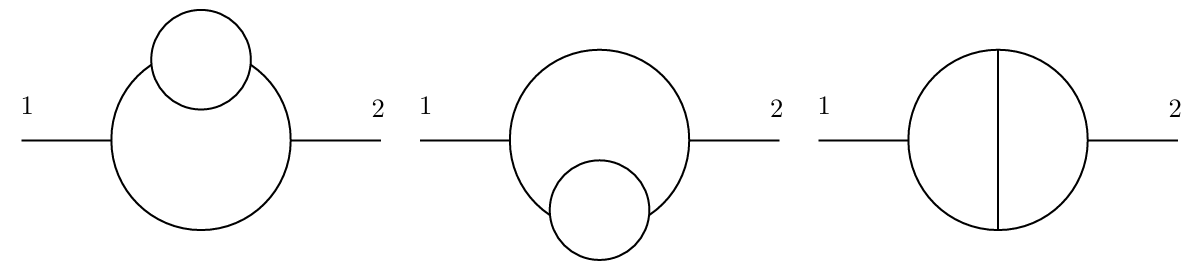}
	\caption{All diagrams that can be generated in the recursion of the 2-way 2-loop kernel. The first two diagrams correspond to the 1-division and the 3-division terms, while the third one corresponds to the 2-division terms. Note that the first two diagrams are actually the same, and this overcounting will be canceled by graph factors.}
	\label{scalar}
\end{figure}

For a 2-way 2-loop kernel polynomial (shown in the figure \ref{scalar}), 1-division terms are
\ie
P^{(1)}=\ma{S}_{a,b}^{(2)}(x^{(1)}_{ab,ee}x^{(1)}_{ba,ee},\{1,2\})+\text{cyclic(1,2)}=x^{(1)}_{ab,ii}x^{(1)}_{ba,ii}x^{(2)}_{12,ee}x^{(2)}_{2b,ei}x^{(2)}_{1a,ei}+x^{(1)}_{ab,ii}x^{(1)}_{ba,ii}x^{(2)}_{12,ee}x^{(2)}_{2a,ei}x^{(2)}_{1b,ei}
\fe
2-division terms are
\ie
P^{(2)}=\ma{S}_{a,b}^{(2)}(x^{(1)}_{ab,ee}x^{(1)}_{bc,ee}x^{(1)}_{ac,ee},\{1\})\bigg|_{c\to2}+\text{cyclic(1,2)}=x^{(1)}_{ab,ii}x^{(1)}_{b2,ie}x^{(1)}_{a2,ie}x^{(2)}_{1a,ei}x^{(2)}_{1b,ei}+x^{(1)}_{ab,ii}x^{(1)}_{b1,ie}x^{(1)}_{a1,ie}x^{(2)}_{2a,ei}x^{(2)}_{2b,ei}
\fe
3-division terms are
\ie
P^{(3)}&=\ma{S}_{a,b}^{(2)}(x^{(1)}_{ab,ee}x^{(1)}_{bc,ee}x^{(1)}_{cd,ee}x^{(1)}_{ad,ee},\emptyset)\bigg|_{c\to1,d\to2}+\text{cyclic(1,2)}\\
&=x^{(1)}_{ab,ii}x^{(1)}_{b1,ie}x^{(1)}_{12,ee}x^{(1)}_{2a,ei}x^{(2)}_{ab,ii}+x^{(1)}_{ab,ii}x^{(1)}_{b2,ie}x^{(1)}_{12,ee}x^{(1)}_{1a,ei}x^{(2)}_{ab,ii}
\fe
In the 1-division terms, there are two 2-way kernel structures for each term. For example, for the monomial $x^{(1)}_{ab,ii}x^{(1)}_{ba,ii}x^{(2)}_{12,ee}x^{(2)}_{2b,ei}x^{(2)}_{1a,ei}$, the 2-way kernel structures are
\ie
(a,\ b):\ x^{(1)}_{ab,ii}x^{(1)}_{ba,ii},\ (1,\ 2):\ x^{(1)}_{ab,ii}x^{(1)}_{ba,ii}x^{(2)}_{12,ee}x^{(2)}_{2b,ei}x^{(2)}_{1a,ei}.
\fe
We can do the same thing for other terms and finally obtain the graph factors for each term. Note that in this case $r=0$ for all terms. For example, again for the monomial $x^{(1)}_{ab,ii}x^{(1)}_{ba,ii}x^{(2)}_{12,ee}x^{(2)}_{2b,ei}x^{(2)}_{1a,ei}$, the largest loop is
\ie
x^{(1)}_{ba,ii}x^{(2)}_{ab,ii}.
\fe
The total contraction of the monomial is
\ie
x^{(1)}_{ab,ii}x^{(1)}_{ba,ii}x^{(2)}_{ab,ii}.
\fe
For each grade, there is only 1 common graded inverse variable between the largest loop and the total contraction of the monomial, and then we have $r=0$. The final result is
\ie\label{22k}
P_{2,2}^{\rm kernel}=\frac{1}{8}P^{(1)}+\frac{1}{4}P^{(2)}+\frac{1}{8}P^{(3)}.
\fe
In \cite{Tao:2025fch}, the graph factor of each term in \eqref{22k} comes from observing the diagrams in figure \ref{scalar}. In our new formalism, we actually do not need to know what the corresponding diagrams are, i.e., we do not need figure \ref{scalar} to help us find out the graph factors. We can figure out the graph factor only by dealing with each monomial.

In general, the bare kernel $P_{\ell,m_{\ell}}^{\rm bare\ kernel}$ with legs $b_1,b_2,\cdots,b_{\ell}$ can be written as the following formalism using the sewing operator
\ie
P_{\ell,m_{\ell}}^{\rm bare\ kernel}=\sum_{m_{\ell-1}=2}^{m_{\ell}+2}\ma{R}\bigg[\ma{S}_{a_1,a_2}^{(\ell)}(P_{\ell-1,m_{\ell-1}}^{\rm bare\ kernel},\{b_1,\cdots,b_{m_\ell-m_{\ell-1}+2}\})\bigg|_{a_i\to b_{m_\ell-m_{\ell-1}+i}\atop 3\leq i\leq m_{\ell-1}}\bigg]
\fe
where we assume the $(\ell-1)$ loop kernel polynomial has legs $a_1,a_2,\cdots, a_{\ell-1}$. Then the whole loop kernel can be obtained from this bare loop kernel by investigating the symmetry factor of each monomial. 

\subsection{The map from monomials to loop kernels}
We have established the recursion relation for the monomials of graded inverse variables. However, what we are really interested in is the integrands. Thus, we need to find an explicit map from monomials to loop kernels. Roughly speaking, the mapping rules are
\begin{enumerate}
    \item For every $x_{kl,ab}$ with $a,\ b=\{e,i\}$, we get the propagator between the vertices $k$ and $l$.
    \item For every $e$ label, we add an extra term $\phi_{k|k}$ where $k$ has the label $e$. For example, for $x_{kl,ie}$ we need to add a $\phi_{l|l}$,
\end{enumerate}
The first mapping rule needs more clarification, which is the goal of this subsection. To simplify the notation, the position labels in the comb component will be ignored in this subsection.

The map from the monomials to loop kernels should be realized in the order of the grade. We first define the map from the comb components:
\ie
{\rm comb}^{(\ell)}(1,2,\cdots,n)\to \frac{1}{(l_{\ell}+k_1)^2(l_{\ell}+k_1+k_2)^2\cdots(l_{\ell}+\sum_{i=1}^{n-1}k_i)^2}
\fe
For the grade-1 part, we have 
\ie\label{grade1map}
x^{(1)}_{12}\cdots x^{(1)}_{(n-1)n}x^{(1)}_{n1}\to\frac{1}{l_1^2(l_1+k_1)^2\cdots(l_1+\sum_{i=1}^{n-1}k_i)^2}
\fe
Then, consider the grade-2 part with a vertex that is common to the lower grade part. There will be two vertices, say $a$ and $b$. Assuming we have sewn a set $(i_1,i_2,\cdots,i_n)$ as the grade-2 part, we will get a factor
\ie\label{factor2}
x^{(2)}_{ai_1}{\rm comb}^{(2)}(i_1,i_2,\cdots,i_n)x^{(2)}_{i_nb}.
\fe
After setting 
\ie\label{grade2map}
k_a\to-l_2,\ k_b\to l_2+\sum_{j=1}^{n}k_{i_j},\ x^{(2)}_{ai_1,ie}\to\frac{1}{l_2^2},\ x^{(2)}_{i_nb,ei}\to\frac{1}{(l_2+\sum_{j=1}^{n}k_{i_j})^2},
\fe
we will get the correct propagators. Specifically, there can also be a factor $x_{ab}^{(2)}$ with both $a$ and $b$ common to the grade-1 vertices. In this case, we have
\ie
k_a\to-l_2,\ k_b\to l_2,\ x_{ab}^{(2)}\to \frac{1}{l_2^2}
\fe
After we have replaced the graded inverse variables with propagators, we need to add the product of some $\phi_{i|i}$s where $i$ represents all the external vertices of a given monomial. 

In general, the grade-$\ell$ factor will be
\ie
x^{(\ell)}_{ai_1}{\rm comb}^{(\ell)}(i_1,i_2,\cdots,i_n)x^{(\ell)}_{i_nb}
\fe
or $x_{ab}^{(\ell)}$ with both $a$ and $b$ common to lower-grade vertices. We have the following map for the $\ell$-grade case:
\ie
k_a\to-l_\ell,\ k_b\to l_\ell+\sum_{j=1}^{n}k_{i_j},\ x^{(\ell)}_{ai_1,ie}\to\frac{1}{l_\ell^2},\ x^{(\ell)}_{i_nb,ei}\to\frac{1}{(l_\ell+\sum_{j=1}^{n}k_{i_j})^2}
\fe
or
\ie
k_a\to-l_\ell,\ k_b\to l_\ell,\ x_{ab}^{(\ell)}\to \frac{1}{l_\ell^2}
\fe
After we have done all the maps grade-by-grade, we will reach the final result, which is expressed as some momenta and $\phi_{i|i}$s.

We give an example here. Consider the monomial with vertices 1 and 2 external:
\ie
x^{(1)}_{ab,ii}x^{(1)}_{ba,ii}x^{(2)}_{12,ee}x^{(2)}_{2b,ei}x^{(2)}_{1a,ei}.
\fe
For grade 1, we have
\ie
\frac{1}{l_1^2(l_1+k_a)^2}
\fe
Then we move to grade 2, we have
\ie
\frac{1}{l_2^2}\frac{1}{(l_2+k_1)^2}\frac{1}{(l_2+k_1+k_2)^2}\phi_{1|1}\phi_{2|2},
\fe
after setting $k_a\to-l_2,\ k_b\to l_2+k_1+k_2$. The final result is
\ie
\frac{1}{l_2^4}\frac{1}{(l_2+k_1)^2}\frac{1}{l_1^2(l_1-l_2)^2}\phi_{1|1}\phi_{2|2}
\fe
using $k_1+k_2=0$. 

To make this map clearer, let us show a further example in detail. Recall the 2-way 2-loop kernel polynomial
\ie
P_{2,2}^{\rm kernel}=&\frac{1}{8}(x^{(1)}_{ab,ii}x^{(1)}_{ba,ii}x^{(2)}_{12,ee}x^{(2)}_{2b,ei}x^{(2)}_{1a,ei}+x^{(1)}_{ab,ii}x^{(1)}_{ba,ii}x^{(2)}_{12,ee}x^{(2)}_{2a,ei}x^{(2)}_{1b,ei})\\
&+\frac{1}{4}(x^{(1)}_{ab,ii}x^{(1)}_{b2,ie}x^{(1)}_{a2,ie}x^{(2)}_{1a,ei}x^{(2)}_{1b,ei}+x^{(1)}_{ab,ii}x^{(1)}_{b1,ie}x^{(1)}_{a1,ie}x^{(2)}_{2a,ei}x^{(2)}_{2b,ei})\\
&+\frac{1}{8}(x^{(1)}_{ab,ii}x^{(1)}_{b1,ie}x^{(1)}_{12,ee}x^{(1)}_{2a,ei}x^{(2)}_{ab,ii}+x^{(1)}_{ab,ii}x^{(1)}_{b2,ie}x^{(1)}_{12,ee}x^{(1)}_{1a,ei}x^{(2)}_{ab,ii})
\fe
After mapping the grade-1 variables using \eqref{grade1map}, we have
\ie
P_{2,2}^{\rm kernel}\to&
\frac{1}{8}(\frac{1}{l_1^2(l_1+k_a)^2}x^{(2)}_{12,ee}x^{(2)}_{2b,ei}x^{(2)}_{1a,ei}+\frac{1}{l_1^2(l_1+k_a)^2}x^{(2)}_{12,ee}x^{(2)}_{2a,ei}x^{(2)}_{1b,ei})\\
&+\frac{1}{4}(\frac{\phi_{2|2}}{l_1^2(l_1+k_a)^2(l_1+k_a+k_b)^2}x^{(2)}_{1a,ei}x^{(2)}_{1b,ei}+\frac{\phi_{1|1}}{l_1^2(l_1+k_a)^2(l_1+k_a+k_b)^2}x^{(2)}_{2a,ei}x^{(2)}_{2b,ei})\\
&+\frac{1}{8}(\frac{\phi_{1|1}\phi_{2|2}}{l_1^2(l_1+k_a)^2(l_1+k_a+k_b)^2(l_1+k_a+k_b+k_1)^2}x^{(2)}_{ab,ii}\\
&+\frac{\phi_{1|1}\phi_{2|2}}{l_1^2(l_1+k_a)^2(l_1+k_a+k_b)^2(l_1+k_a+k_b+k_2)^2}x^{(2)}_{ab,ii})
\fe
Then, referring to \eqref{factor2} and \eqref{grade2map}, after using $k_1+k_2=0$ we have
\ie
P_{2,2}^{\rm kernel}\to&
\frac{1}{8}(\frac{\phi_{1|1}\phi_{2|2}}{l_1^2(l_1-l_2)^2l_2^4(l_2+k_1)^2}+\frac{\phi_{1|1}\phi_{2|2}}{l_1^2(l_1+l_2)^2l_2^4(l_2+k_1)^2)}\\
&+\frac{1}{4}(\frac{\phi_{1|1}\phi_{2|2}}{l_1^2(l_1-l_2)^2(l_1+k_1)^2l_2^2(l_2+k_1)^2}+\frac{\phi_{1|1}\phi_{2|2}}{l_1^2(l_1-l_2)^2(l_1+k_2)^2l_2^2(l_2+k_2)^2})\\
&+\frac{1}{8}(\frac{\phi_{1|1}\phi_{2|2}}{l_1^4(l_1-l_2)^2(l_1+k_1)^2l_2^2}+\frac{\phi_{1|1}\phi_{2|2}}{l_1^4(l_1-l_2)^2(l_1+k_2)^2l_2^2})
\fe
After shifting the loop momenta properly, the final result can be written as
\ie
P_{2,2}^{\rm kernel}\to\frac{\phi_{1|1}\phi_{2|2}}{2l_1^2(l_1-l_2)^2l_2^4(l_2+k_2)^2}+\frac{\phi_{1|1}\phi_{2|2}}{2l_1^2(l_1-l_2)^2(l_1+k_1)^2l_2^2(l_2+k_1)^2},
\fe
which is the same as the result in \cite{Tao:2025fch}.

\section{Conclusion}
In this paper, we showed a more elegant formalism to express the loop kernel recursion using graded inverse variables. Our results showed that there are still many interesting mathematical structures and techniques that need to be found and developed.

It is worth generalizing the graded inverse algebra to more general theories, such as the Yang-Mills theory. We can also try to find some differential equations involving these variables, as an analog of \cite{Arkani-Hamed:2024pzc}. It is also interesting to see the connection between the structure here and the algebraic structure of the perturbiner expansion \cite{Lopez-Arcos:2019hvg}.

\section*{Acknowledgements}
YT thank Chen Huang for some suggestions on the draft. YT is supported by the National Key R\&D Program of China (NO. 2020YFA0713000 and NO. 124B2094).

\bibliographystyle{JHEP}
\bibliography{planarloop}

\end{document}